# THE OPTIMIZATION OF RUNNING QUERIES IN RELATIONAL DATABASES USING ANT-COLONY ALGORITHM

Adel Alinezhad Kolaei and Marzieh Ahmadzadeh

Department of Computer Engineering & IT Shiraz University of Technology

*ABSTRACT*

*The issue of optimizing queries is a cost-sensitive process and with respect to the number of associated tables in a query, its number of permutations grows exponentially. On one hand, in comparison with other operators in relational database, join operator is the most difficult and complicated one in terms of optimization for reducing its runtime. Accordingly, various algorithms have so far been proposed to solve this problem. On the other hand, the success of any database management system (DBMS) means exploiting the query model. In the current paper, the heuristic ant algorithm has been proposed to solve this problem and improve the runtime of join operation. Experiments and observed results reveal the efficiency of this algorithm compared to its similar algorithms.*

*KEYWORDS*

*Database, Join Operation Optimization , Traveling Salesperson, Ants Algorithm.*

## 1. INTRODUCTION

Optimization of database queries such as relational database queries is one of the most important research issues. If queries are in interactional mode, they will contain a few relations such that optimization of these queries can be done using a comprehensive search. However, if the number of relations is more than 5 or 6, techniques for comprehensive search will be costly in terms of computer memory and time. Optimization of queries is an activity in which an efficient strategy for performing the query is produced hence it is a fundamental step in query processing. In this stage, DBMS selects the best strategy among some given executive strategies so that running the given query using this strategy by user has minimum cost. The input in this system is a query such as q to DBMS by user. Suppose that S is the set of all possible strategies to answer query q. Each element of the set S such as s has cost $C(s)$ (in terms of CPU usage, input and output, memory consumption, etc.). The purpose of an optimization algorithm to find an element $s_0$ in S such that [1,2]:

$C(s_0) = \min C(s_i)$ si ε S

    

In order to respond to query q, a strategy applies a sequence of algebraic operators on database relations so that it ultimately finds the answer of q. The overall cost of a strategy is the sum of processing cost of each operator. In comparison to existing relational operators, processing and optimizing join operator which is denoted by $\infty$ is one of the most complicated tasks. Basically, the join operator uses two relations as input and combines tuples of each relation correspondingly and returns a new relation as the output. Since join operator has commutative and associative properties, the number of existing running strategies to answer a query grows exponentially as the number of joins between relations increases; however, all of the existing running strategies to answer to a specified query have not the same output (such as $R_3=R_1 \infty (R_2 \infty R_3) \infty (R_1 \infty R_2)$), the resultant running strategy has a different cost, since the cardinality of alternative relations are not the same. So, selecting an appropriate ordering for running join operator is influential to the overall cost. Since, optimization of a query which consists of join operator is more time-consuming than other relational operators such as selection or projection operator, all of the query optimization techniques are applied to queries involved in join operator. Some algorithms which can be used to solve this problem are as follow:

Iterative Improvement, Genetic Algorithm, Particle Swarm Optimization, Simulated Annealing, Hybrid Evolutionary Algorithm, etc. It can be understood that hyper-heuristic algorithms have been resulted in appropriate solutions to this problem. Accordingly, in the present paper, we have used hyper-heuristic ant algorithm to solve this problem.

The contents of this paper have been organized as follows:

In sections 2 and 3, Traveling sales person problem and a brief history of ant Algorithm is respectively mentioned. In section 4, the proposed Algorithm is illustrated. In sections 5 and 6, implementation environment and results of experiments are explained, respectively. In section 7, the paper is concluded.

## 2. TRAVELING SALES PERSON PROBLEM

Traveling Sales Person problem is a classic and famous problem in the field of operation research. Many problems in science can be formulated in the form of this problem and then solved. This problem consists of n cities where there will be a path between each two cities. Every such path between two cities has a cost function which equals to the distance between these cities. Traveling sales person wants to start his travel from one of the cities and then visits all of the cities just once and finally returns to the first city where he started his travel. The solution for this problem is a sequence of cities where the sales person has visited once and the main constrain is that the total cost (distance) be minimized [7].

## 3. ANT ALGORITHM

In 1991, Dorigo and Clarini used hybrid-heuristic ant Algorithms to solve separable problems [8]. This Algorithm has such properties as simple adaptability for applying in many problems. Ant Algorithms are multi-agent systems whose behavior is derived from real ants and each agent is an artificial ant. These Algorithms are successful samples of swarm intelligence systems and are applicable to a range of problems from Traveling Sales Person problem to routing in remote communication networks. In addition, they can be used in hybrid optimization problems such as quadratic allocation and scheduling task problems in a set of given tasks.

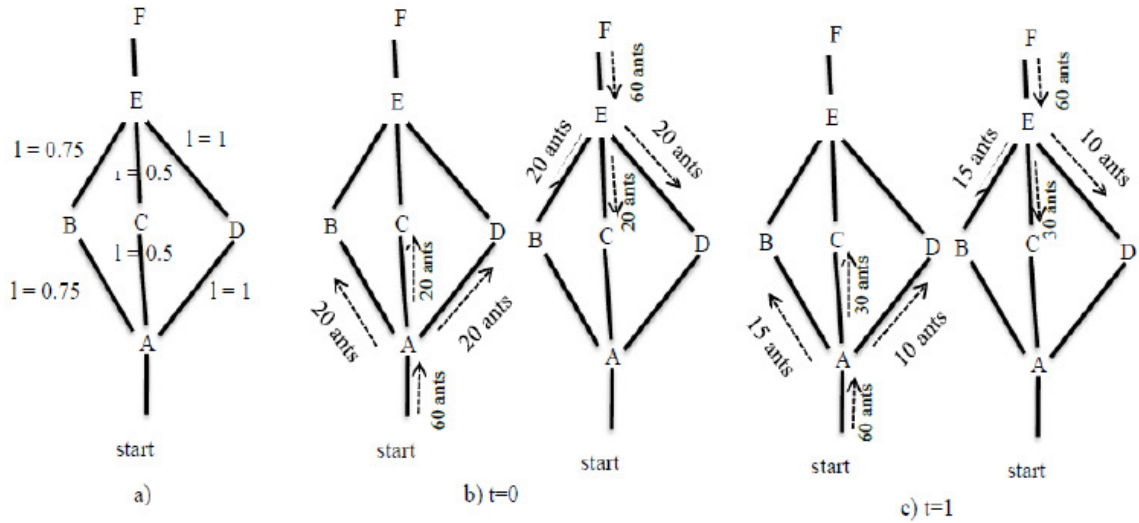

Figure 1. The way of moving ants to find food resources and shedding pheromone

Figure (1) shows the way that ants adopt to find food resources. When ants move they remain a chemical substance called pheromone. As it is shown in figure (1), at time t=0 the number of ants in two different paths are the same but at time t=1 their numbers in these paths are different and in shorter path the number of ants is more. This is the case because during their movement, ants release pheromone and as time passes, it is evaporated. Therefore, the shorter the length of the path, the less the pheromone is evaporated and the more the remaining pheromone will be, so ants will select this shorter path through next stages.

The experiment on ants shows that despite the existence of two paths with different length from nest to food, after a while (about a few minutes), most of ants select shorter path and this ratio increases as the difference between paths increases. In order to find food, ants release pheromone while moving from nest to other place and vice versa. When they reach a point which is the intersection point of shorter and longer path, they adopt a probabilistic selection based on the amount of pheromone smelled. Such behavior has an autocatalytic effect because when a path is selected, it is more probable that in future the same path will be selected too. Therefore, by repeating this procedure, the released pheromone in the shorter path will be saved in higher rate so that the shorter path is more and more selected. This simple idea is applicable to find appropriate solutions in complicated optimization problems [8, 9, 10, 11].

In Traveling Sales Person problem, there are a set of n cities in which we should find the shortest closed path consisting of these cities. If dij denotes the length of path between cities i and j in symmetric Traveling Sales Person problem, to find the shortest path among these cities we will use some artificial ants (agents) which have the following properties:

Ants will randomly be placed on nodes.

Each ant selects the next city in a probabilistic approach. The probability of selecting next city is a function of distance to other cities the associated value in the sequence of cities on the related nodes.

Each ant has to move to allowable paths i.e. it should move to cities that was not visited in the past. This condition is controlled using prohibited list.

When a path is formed, pheromone is placed on its nodes and the ant releases some pheromone while crossing the cities. Ant in the time interval (t, t+1) performs one move, so, in each iteration of Algorithm, m move will occur.

In the time interval (t, t+1), some pheromone evaporates. When each ant has created its own path, pheromone will again be specified to them [8, 9, 10, 11].

In 1996, Dorigo and Jembardla, made some modifications to ant Algorithm [9] and proposed Ant-Colony Algorithm. This proposed Algorithm was different from previous Algorithm in terms of following characteristics:

- The changing position rule which directly controls the effect of new nodes and old ones on the Algorithm.
- The overall updating rule that updating had been used only on nodes which belonged to ants' paths.
- When the ants form a solution, the local updating rule will be applied.

## 4. OUR PROPOSED APPROACH

Considering the proposed illustrations, it can be stated that if we can formulate optimization problems in well-known problems for which some Algorithms have been proposed, we can solve them through the above-mentioned methods. Therefore, the main task is how to convert these problems in to famous solved problems. As it was mentioned above, one of the famous problems is Traveling Sales Person problem for which different Algorithms have been proposed to solve it. Such Algorithms are Genetic Algorithm, Particle Swarm Optimization, Ant Algorithm, hybrid Algorithm, etc.

We decide to solve optimizing join operation problem in relational database firstly by converting it to TSP problem and then solve it using Hyper-Heuristic Ant Algorithm. It should be mentioned that the main problem in this paper is not the same to original TSP problem because in the original TSP problem, sales person returns to first city, however, in the proposed problem, this is not the case. So, firstly formulas and the way that this problem should be solved using ant Algorithm has been stated and then important parameters and how converting this problem will be stated.

Ant-Colony Algorithm which we considered in this paper is exactly the one proposed by Dorigo [9]. In this Algorithm, in the first phase, initially m ants with memory are created. These ants will randomly be placed on n nodes. On each node, there is some initial pheromone. In the second phase, to obtain initial solutions, the following steps are run in parallel.

To select city s in its next move, the ant which is placed on node r, uses relation (1) known as position changing rule in Ant-Colony Algorithm. In relation (1), q is a random number in [0, 1] interval and the value of q0 is between 0 and 1.

$$s = \begin{cases} \arg\max_{u \in J_k(r)} \{[\tau(r,u)].[\eta(r,u)]^\beta\} & if \quad q \leq q_0 \\ S & otherwise \end{cases} \quad (1)$$

Where τ(r, s) represents the amount of pheromone on arc (r,s), η(r, s) is the reciprocal distance δ(r,s), JK(r) is the remaining set of cities for k-th ant placed on city r, β is the importance determining parameter in the relation between pheromone and distance. However, in relation (1), s is a random variable that follows from the probability distribution mentioned as relation (2).

$$s = \begin{cases} \dfrac{[\tau(r,s)].[\eta(r,s)]^\beta}{\sum_{u \in J_k(r)} [\tau(r,u)].[\eta(r,u)]^\beta} & if \quad s \in J_k(r) \\ 0 & otherwise \end{cases} \quad (2)$$

Where Pk(r,s) is the probability that k-th ant after selecting city r will select city s. After that ants create their own paths, they return to their first city. In this step, local updating will be done and the amount of pheromone changes by relation (3).

$$\tau(r,s) \leftarrow (1-\rho).\tau(r,s) + \rho.\Delta\tau(r,s) \quad (3)$$

Where ρ is pheromone evaporating parameter and Δτ(r,s) is obtained using relation (4):

$$\Delta\tau(r,s) = \begin{cases} (L_{gb})^{-1} & if \quad (r,s) \in global\_best\_tour \\ 0 & otherwise \end{cases} \quad (4)$$

where Lgb is the length of the best path which has already been found in the current iteration [9]. To solve the join problem of tables using Ant-Colony Algorithm, it suffices that we consider every table as a city (n cities) and that sales person doesn't return to first city and also we should consider η(r, u) equal to reciprocal of the cost of two tables r and u and Lgb as minimum cost of the join of all tables in each iteration of Algorithm (parameter β can also be considered as an important parameter in distributed database). Therefore, in beginning of Algorithm, each ant is randomly placed on a city, then for move, join of this city with other cities (tables) should be verified if joining of these cities is possible as mentioned in the definition of problem-that is there exists a path between them- and after these steps, the table with minimum cost (least time for join operation) will be considered as a next city (table).

To achieve this purpose, in each step, the cost of joining two tables as a cost of path between these two cities is added up with the cost of previous joins (equals total cost up to current step). In each run of Algorithm, with respect to local updating, the amount of pheromone changes until the best path is specified, so this path can be used in the next step and finally the best path (the join with minimum cost) can be selected.

## 5. IMPLEMENTATION ENVIRONMENT

We have used databases with the number of tables equal to 8, 12, 16, 20, 24, 28 and 32 in which the number of records in each table has been produced as a random number with uniform distribution. The parameters and system specification and implementation environment are as follows:

We have used a system with Intel Core 2duo and 2.26GHZ cpu, 3GB RAM memory XP operating system. The language that program has been written with is C# and also we have used SQL SERVER database. In addition, we ran the program 20 times and in each run we have used 30 iterations, in other words, Iteration Number=30 and Run Number=20. The mean number of records which are used from each table in experiments is given in table1:

Table1 : The mean number of records in each table in experiments

| Number of tables | Average number of records per table |
|---|---|
| 8 | 383 |
| 12 | 418 |
| 16 | 362 |
| 20 | 397 |
| 24 | 403 |
| 28 | 354 |
| 32 | 429 |

## 6. EXPERIMENTAL RESULTS

For our proposed approach, the number of ants is considered as the number of vertices and then the program has been executed. The number of iteration s for each experiment was 20 and the mean number of times through which iterations have been done is given in figure2.

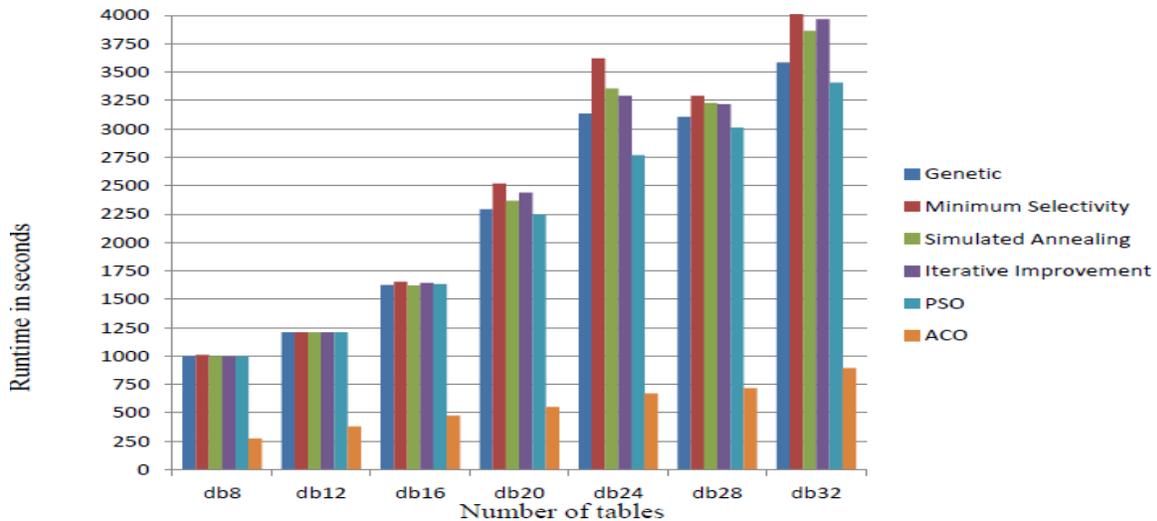

Figure 2. Execution time tests

As it can be seen from figure 2, the proposed approach has resulted in a good performance in comparison to other approaches. This is the case because other approaches verify almost all existing ways of running a query while in the proposed approach , in each experiment if ants find a better solution (to join tables), they will inform the existence of this solution with respect to the updating strategy to other ants in order that they use this solution to find optimal solutions. Therefore, all existed solutions for running a query (join of multiple tables) will have not been verified and  cpu time  is saved. In other words, in our approach, ants operate in parallel.

It should be noted that in the proposed approach, we have set the number of ants equal to the number of tables w and in the next time, we set it as half the number of tables and finally we have shown that for the second case, cpu time has decreased. On the other hand, the more the number of ants is, the less the decrease in the probability of the precision of the obtained result (as the experiment shows). Therefore, if we use more powerful processors, the runtime of Algorithm will decrease. In other words, if we use multi-processor system, runtime will decrease to an acceptable level and this is the case because ants Algorithm operates parallel and also the ants work in parallel with each other and updating is performed when each step is finished.

## 7. CONCLUSION

In this paper, we stated a new approach for optimizing execution of queries in relational databases using Ant Colony algorithm. For this, we initially maped the  connection between tables into a TSP model ,in which   each table was considered as a city and each  table link as a relation between two cities. Then the problem was solved by Ant Colony algorithm. Several such experiment was   conducted for a number of different tables for which the numbers of records in each table were chosen randomly with uniform distribution. Experiments and given results proved the better performance of this approach rather than others that were given in the literature.  The main reason for the efficiency of this method is its parallel property. This happened since our approach did not investigate all the existing ineffective and useless solutions for query execution, while, in the majority of other approaches the most existing solutions need to be evaluated. . As a consequence, we achieved a better execution time.

## AUTHORS

Adel Alinezhad Kolaei is a MS student of Computer Science at Shiraz University of Technology. He received his BSc in Software Engineering from Iran University of Science and Technology. His major research interests are high performance algorithms and database management.

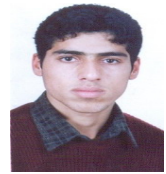

Marzieh Ahmadzadeh is an Assistant Professor of Computer Science at Shiraz University of Technology. She received her PhD in Computer Science and MSc in Information Technology from the University of Nottingham, UK and her BSc in Software Engineering from Isfahan University, Iran. She teaches a variety of graduate and undergraduate courses and her research interest includes Computer Science Education in general and Computer Supported Learning, specifically Data Mining and Human Computer Interaction.

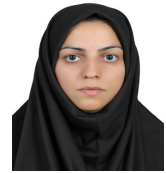